\documentstyle[12pt]{article}

\def\bc{\begin{center}}                \def\ec{\end{center}}
\def\be{\begin{equation}}              \def\ee{\end{equation}}
\def\bear{\begin{eqnarray}}            \def\eear{\end{eqnarray}}
\def\bt{\begin{tabular}}               \def\et{\end{tabular}}
\def\la{\langle}     \def\ra{\rangle}   \def\dg{\dagger}
\def\ci{\cite}       \def\lb{\label}     \def\ld{\ldots}
\def\hs{\hspace}      \def\vs{\vspace}   \def\pr{\prime}
\def\sm{\small}        \def\p{\partial}   \def\nn{\nonumber}
  
\def\td{\tilde}       \def\pr{\prime}

\def\a{\alpha}      \def\b{\beta}    \def\g{\gamma}     \def\G{\Gamma}
\def\D{\Delta}     \def\d{\delta}    \def\e{\epsilon}   \def\k{\kappa}
\def\l{\lambda}    \def\L{\Lambda}    \def\s{\sigma}    \def\z{\zeta}
\def\o{\omega}   \def\O{\Omega}       \def\r{\rho}      \def\t{\theta}
      \def\Re{{\rm Re}} \def\Im{{\rm Im}}

\normalbaselines
\begin{document}
\begin{flushright}                             quant-ph/9811081  \\
                                  J. Phys. A 32(19) (1999) 3649-61
\end{flushright}
\vspace{8mm}

\bc
{\large \bf    Exact Solutions for the  General Nonstationary
             Oscillator with a Singular Perturbation }\\[3mm]

                    {\bf  D.A. Trifonov }\\[1mm]

{\it Institute for Nuclear Research and Nuclear Energetics\\
            72 Tzarigradsko chauss\'ee, Sofia 1784, Bulgaria }
\ec
\vs{5mm}

{\bf Abstract}

{\small
Three linearly independent Hermitian invariants  for the nonstationary
generalized  singular oscillator (SO) are constructed and their complex
linear combination is diagonalized. The constructed family of eigenstates
contains as subsets all previously obtained solutions for the SO and
includes all Robertson and Schr\"odinger intelligent states for the three
invariants. It is shown that the constructed analogues of the $SU(1,1)$
group-related coherent states for the SO minimize the Robertson and
Schr\"odinger uncertainty relations for the three invariants and for every
pair of them simultaneously.  The squeezing properties of the new states 
are briefly discussed.} 
\vs{1cm}

\section{\large \bf Introduction}

Recently  a considerable attention has been paid in the literature
\ci{DMR,Um,Kaushal,Simon,DMZ} to the (nonstationary) singular
oscillator, i.e. the particle with mass $m$ in the harmonic plus an
inverse harmonic potential\\[-7mm]

\be\lb{V}                                   
V(x) = \frac{1}{2}m\o^2x^2 + g\frac{1}{x^2},
\ee
where the mass and/or frequency may depend on time. Previously this
singular oscillator (SO) has been treated in a number of paper
\ci{Cheng,CDM,Korsch,THM81,Ray,DMM,Calogero,Gamiz,Peak}, exact invariants
and wave functions being obtained for the case of stationary SO (constant
$m$, $\o$ and $g$) in \ci{Calogero,Peak} and of SO with varying frequency
$\o(t)$ (but constant $m$, $g$) in \ci{DMM, Gamiz}. The more general cases
treated in \ci{CDM,THM81} corresponds (as in \ci{Um}) to $m(t)g(t) ={\rm
const}$.  The potential (\ref{V}) has vide applications in molecular and
solid state physics: the radial motion of such systems as the hydrogen atom,
the $n$-dimensional oscillator, the charged particle in an uniform magnetic
plus electric field with scalar potential proportional to $1/(x^2+y^2)$ and
the $N$ identical particles interacting pairwise with potential energy
$V_{ij} = (m\o^2/2)(x_i-x_j)^2 + g/(x_i-x_j)^2$ can be reduced
\ci{Calogero,Gamiz,BR} to the case of SO.  The potential (\ref{V}) was
recently applied to describe a two-ion trap \ci{DMR}.

The aim of the this paper is to construct new family of exact wavefunctions
for the SO and to extend these and the previous solutions
\ci{Calogero,DMM,THM81,Um} to the case of nonstationary general oscillator
with the singular perturbation $g(t)/x^2$,\,\, $m(t)g(t) = {\rm const}$
(general SO). To this aim we make an efficient use of the method of
time-dependent quantum invariants \ci{Lewis,MMT1,MMT2}.  We construct a
four complex parameters ($z,u,v,w$) family of states $|z,u,v,w;\k,t\ra$ of
general SO, which diagonalize the general complex combination of the three
linearly independent Hermitian invariants $I_j(t)$.

The large family $|z,u,v,w;\k,t\ra$ of general SO states contains all
previously obtained solutions as subsets and new states with interesting
properties.  In particular it contains the analogues of the
Barut-Girardello coherent states (CS) and the $SU(1,1)$ group-related CS
with symmetry \ci{BR,KS}. The most important physical properties of the
new wavefunctions of the SO (which the previous solutions lack) are the
strong squeezing \ci{Walls,Trif94} in the $SU(1,1)$ generators and the
maximal intelligency \ci{Aragone,Trif94} with respect to Robertson
\ci{Rob} and Schr\"odinger \ci{Rob} uncertainty relations.  In particular
the new states can exhibit strong quadratic squeezing \ci{Hillery}.  The
obtained family of states can be regarded as an extension of the
$su^C(1,1)$ algebra-related coherent states (CS) $|z,u,v,w;k\ra$
\ci{Trif94,Trif9698,Brif} from the series $D^+(k)$ with discrete Bargman
index $k=1/4,3/4$, $k=1/2,1,\ldots$ \ci{BR} to the series with continuous
Bargman index $\k \geq 1/2$.  Recall that the nonstationary quadratic in
the coordinate and the momentum Hamiltonian, which has wide applications
in quantum optics (see \ci{Walls} and references therein), is an operator
from the $su(1,1)$ representation with Bargman index $k = 1/4,3/4$.

In section 2 the three independent invariants $I_j(t)$ for the general SO
are constructed and expressed in terms of a complex parameter $\e(t)$ which
obey the classical oscillator equation.  The three invariants close the
$su(1,1)$ algebra in a continuous series representations.  In section 3 the
general complex combination of the invariants $I_j(t)$ is diagonalized and
some limiting cases of the corresponding eigenstates $|z,u,v,w;\k,t\ra$, are
considered. The Green function of the general SO is also written down. The
overcompleteness, intelligent and the squeezing properties
\ci{Walls,Aragone,Trif94} of the constructed states are briefly outlined.
It is shown that the wavefunctions $\Psi_\xi(x,t)$, which are the analogue
of the $SU(1,1)$ group-related CS with symmetry, are states with maximal
intelligency, i.e. they minimize simultaneously the Schr\"odinger
uncertainty relation for every three pairs of invariants $I_i(t),\,I_j(t)$
and the Robertson inequality for the three invariants.  Finally, in section
4 we give a summary and some concluding remarks.

\section{Symmetry and invariants for the general SO}

The Hamiltonian of the general oscillator with a singular perturbation
we consider is of the form \\[-5mm]

\be\lb{H}                                    
H(t) = \frac{1}{2m(t)}p^2 + b(t)(px+xp)
        +\frac{m(t)\o^2(t)}{2}x^2 + \frac{g(t)}{x^2},
\ee
where $m(t),\, \o(t)$ and $b(t)$  are arbitrary real differentiable
functions, $m(t)>0$. The time-dependence of $g(t)$ is related to that of
$m(t)$ as

\be\lb{c}                                   
2m(t)g(t)/\hbar^2 = c = {\rm const},
\ee
where $c$ is arbitrary real (dimensionless) parameter. As we shall see
below the constrain (\ref{c}) ensures $H(t)$ belongs
                                                  
to the algebra $su(1,1)$.  For brevity system (\ref{H}) should be
referred to as the general SO. Exact invariants and wave functions for
various particular cases of (\ref{H}) have been considered in the
literature:  $b(t) = 0$, constant $m$, $\o$ and $g$ -- in
\ci{Calogero,Peak}; $b(t) = 0$, constant $m$, $g$ and varying $\o(t)$ --
in \ci{DMM,Gamiz}; $b(t) = 0$, varying $m(t)$, $\o(t)$, $g(t)$ with the
constrain (\ref{c}) with $c=1$ -- in \ci{Um}; varying $b(t)$ and $\o(t)$
and constant $m$ and $g$ -- in \ci{THM81}.  In \ci{CDM} three Heisenberg
operators and their correlation functions for (\ref{H}) with (\ref{c})
were considered (in different parametrization). In this section we
construct three linearly independent exact invariants $I_j(t)$ for the
general SO (\ref{H}), the invariants being expressed in terms of a complex
time-dependent parameter $\e(t)$ which obey the classical harmonic
oscillator equation. The invariants $I_j(t)$ are represented as
time-dependent linear combinations of the three $su(1,1)$ operators $L_j$.

It is worth noting that the time dependence of the mass (and of the coupling
$g(t)$ if (\ref{c}) is valid) can be eliminated by the simple time-scale
transformation, $$t \rightarrow t^\pr = \int^t d\tau\,m_0/m(\tau).$$ The
resulting Hamiltonian $H^\pr$ is with constant mass $m_0$ and coupling $g_0
= mg/m_0$ and new time-dependent frequency $\o^\pr = m\o/m_0$ and $b^\pr =
mb/m_0$.  For the latter Hamiltonian exact invariants, wave functions and
Green function were constructed in
\ci{THM81}.

The operator $px+xp$ is easily recognized as the pure squeezing generator
\ci{Walls}.  One can check that the time-dependent canonical
transformation, generated by the squeeze operator $S(\td{b}(t)) =
\exp[(-i/\hbar)\td{b}(t)\,(xp+px)]$, $\td{b}(t) = \int^tb(\tau)d\tau$,
converts (\ref{H}) into SO Hamiltonian $H^\pr$ with the same frequency and
new mass $m^\pr= m\exp[-4\td{b}(t)]$ and new coupling $g^\pr = g
\exp[4\td{b}(t)]$. Time-dependent canonical transformations are in fact
very powerful -- they can convert a given Hamiltonian into any desired one
\ci{Trif98}. On the classical level switching on the term $b(xp+px)$
results in a sudden change of the squared frequency from $\o^2$ to
$\o^2-\o^2_1(b)$, $\o^2_1 = 4b^2+2\dot{b}+2b\dot{m}/m$.  If $\o^2_1 >
\o^2$ one gets the inverted oscillator. This motivates the term 'general
oscillator' for the Hamiltonian system (\ref{H}) with $g=0$.

We first construct the time-dependent invariants for the system
(\ref{H}). The defining equation of the invariant operators $I(t)$ for a
quantum system with Hamiltonian $H$ is \\[-5mm]

\be\lb{I}                                    
\frac{dI(t)}{dt} = \frac{\partial I(t)}{\partial t} -
\frac{i}{\hbar}[I(t),H] = 0.
\ee
   Formal solutions to this equation  are operators $I(t) =
U(t)I(0)U^\dg(t)$, where $U(t)$ is the evolution operator of the system,
$U(t) = T\exp[-(i/\hbar)\int^tH(\tau)d\tau)]$.  However the explicit
construction greatly simplifies if one can guess the operator structure
of $I(t)$ and substitute it in (\ref{I}). The method of time-dependent
invariants was developed and efficiently used in \ci{Lewis,MMT1,MMT2} to
construct exact wavefunctions for the varying frequency and mass oscillator
\ci{Lewis,MMT1} and for $n$-dimensional nonstationary quadratic systems
\ci{MMT2}. In particular the time evolution of the Glauber CS and Fock
states was explicitly found for general quadratic Hamiltonians \ci{MMT2}
[Quantum mechanical studies of quadratic Hamiltonians can be found in many
later papers (see \ci{Yeon} and references therein)].  In paper \ci{CDM}
three linearly independent Heisenberg operators for the general SO were
constructed in the form of elements of the $su(1,1)$ algebra.

We are looking for solutions of eq. (\ref{I}) for general SO
(\ref{H}) of the same  operator form as that of the Hamiltonian,\\[-5mm]

\be\lb{Igso}                               
I(t) = \a(t)p^2 + \b(t)(xp+px) + \g(t)x^2 + \d(t)\frac{1}{x^2}
\ee
where $\a(t),\,\b(t),\, \g(t)$ and $\d(t)$ may be real or complex
functions of time (then $I(t)$ would be Hermitian or non-Hermitian
invariant). Substituting (\ref{Igso}) and (\ref{H}) into (\ref{I}) (and
assuming $[x,p]=i\hbar$) we obtain equations for the above coefficients
\\[-5mm]

\be\lb{cond1}                              
\a(t) g(t) = \d(t)/2m(t),
\ee
\be\lb{cond2}                              
\left.\begin{tabular}{l}
$\dot{\a}(t) = 4\a(t)b(t)-2\b(t)/m(t)$, \\
$\dot{\b}(t) = 2\a(t)m(t)\o^2(t)-\g(t)/m(t)$,  \\
$\dot{\g}(t) = 2\left[\b(t)m(t)\o^2(t)-2\g(t)b(t)\right]$,  \\
$\dot{\d}(t) = 4\left[\d(t)b(t)-\b(t)g(t)\right]$.
\end{tabular}\right\}
\ee

From (\ref{cond1}) and the first and the fourth equations in (\ref{cond2})
we easily obtain the constrain (\ref{c}) on the time evolution of $m(t)$
and $g(t)$.  Thus for the nonstationary general SO invariants of the form
(\ref{Igso}) exist only under constrain (\ref{c}). The Lie
algebraic meaning of the latter is that it is under this constrain only
when the Hamiltonian of general SO is an operator of the algebra
$su(1,1)$.  Indeed, the  four  operators $p^2$, $xp+px$, $x^2$ and $1/x^2$
do not close any algebra  under commutations, but the three combinations
$L_i$, \\[-5mm]

\be\lb{L_j}                               
L_1 = \frac 14(Q^2-P^2 -\frac{c}{Q^2}),\hs{5mm}
L_2 = -\frac 14(QP + PQ),\hs{5mm}
L_3 = \frac 14(Q^2+P^2 +\frac{c}{Q^2}),
\ee
where $c$ is arbitrary real constant, and $Q$ and $P$ are dimensionless
coordinate and moment,

$$Q = x\sqrt{m_0\o_0/\hbar},\quad P = p/\sqrt{m_0\o_0\hbar},$$
close the algebra $su(1,1)$ \ci{BR},

$$[L_1,L_2] = -iL_3,\,\, [L_2,L_3] = iL_1,\,\,[L_3,L_1]=iL_2.$$
In the above $m_0$ and $\o_0$ are  parameters with the dimension of mass
and frequency respectively. They  may be treated as the initial values of
$m(t)$ and $\o(t)$ in (\ref{H}).  One can easily check that $H(t)$,
equation (\ref{H}), is a linear combination of $L_j$ if and only if the
constrain (\ref{c}) is satisfied:  \\[-5mm]

\be\lb{H2}                               
 H(t) =
\hbar\o_0\left[\left(\frac{m(t)\o^2(t)}{m_0\o^2_0} -
          \frac{m_0}{m(t)}\right)\,L_1 -
          4\frac{b(t)}{\o_0}L_2 + \left(\frac{m(t)\o^2(t)}{m_0\o^2_0}
          +\frac{m_0}{m(t)}\right)\,L_3\right] \equiv h_j(t)L_j.
\ee
This $H(t)$ would be the general element of the algebra $su(1,1)$ if
$h_j(t)$ can acquire arbitrary real values. This can be achieved if $\o^2$
can take any real values, not only positive ones (in order for $h_3$ to be
arbitrary real), i.e.  if the nonsingular part of $H(t)$ is the general
quadratic in $p,\,q$ form, including the inverted oscillator. In view of the
above symmetry the general SO described by (\ref{H}) and (\ref{c}) can be
adequately called (general) {\it $su(1,1)$ SO}.

The Casimir invariant of the algebra spanned by $L_j$ is \\[-5mm]

\be\lb{C_2}                              
C_2 = L_3^2 - L_1^2 - L_2^2 = -\frac{3}{16} + \frac{c}{4} = \k(\k-1),
\ee
where $\k =1/2 \pm (1/4)\sqrt{1+4c}$.  The relation (\ref{C_2}) means that
the representation realized by $L_j$ is reducible and the dynamical
symmetry group \ci{MMT2,BR} of the nonstationary general SO with a fixed
value of $c$ in (\ref{c}) is $SU(1,1)$. Note that for a given
$c\equiv 2mg/\hbar^2$ there are two different values of the  parameter
kappa except for the case of $c=-1/4$. Kappa is real for $c \geq -1/4$ and
complex for $c<-1/4$.  The SO solutions of the previous publications
\ci{CDM}--\ci{Peak} are expressed in terms of $g$ or $a =
(1/2)\sqrt{1+8mg/\hbar^2}$. We find the continuous parameter $\k$ (to be
called the Bargman parameter) most convenient with regards to the maximal
analogy with the solutions related to the discrete series representations
$D^+(k)$ of $su(1,1)$.

The three {\it linearly independent invariants} $I_j(t)$ for the general
SO under the constrain (\ref{c}) are found in the form of the
following time-dependent linear combinations of the dimensionless
$su(1,1)$ operators $L_j$\\[-5mm]

\bear                                   
I_1(t) =
 2\hbar\left[\left(\frac{1}{m_0\o_0}\Re\,\g(t)-
m_0\o_0\Re\,\a(t)\right)\,L_1 - 2\Re\,\b(t)\,L_2\right]\qquad \nn \\
+ 2\hbar\left(
\frac{1}{m_0\o_0}\Re\,\g(t) + m_0\o_0\Re\,\a(t)\right)\,L_3\, , \quad
\lb{I_1}\\                                   
I_2(t) =
 2\hbar\left[\left(-\frac{1}{m_0\o_0}\Im\,\g(t)+
m_0\o_0\Im\,\a(t)\right)\,L_1 + 2\Im\,\b(t)\,L_2\right]\qquad \nn \\
- 2\hbar\left(
\frac{1}{m_0\o_0}\Im\,\g(t) + m_0\o_0\Im\,\a(t)\right)\,L_3\, , \quad
\lb{I_2}\\                                              
I_3(t) =
8\hbar^2\left[ \left(-m_0\o_0\Im(\a(t)\b^*(t)) +
 \frac{1}{m_0\o_0}\Im(\g(t)\b^*(t))\right)\, L_1 - {\rm
Im}(\a(t)\g^*(t))\,L_2\right]\quad \nn \\
+ 8\hbar^2\left(m_0\o_0\Im(\a(t)\b^*(t))
 + \frac{1}{m_0\o_0}\Im(\g(t)\b^*(t))\right)\, L_3\, ,\quad
\lb{I_3}                                                 
\eear
where $c$ is arbitrary constant and $\a(t),\,\b(t)$ and $\g(t)$ are
expressed in terms of one complex parameter $\e(t)$ which obey the
classical oscillator equation, \\[-5mm]

\bear\lb{epsilon}
\ddot{\e}(t) + \O^2(t)\e(t) = 0,\qquad\\            
\O^2(t) = \o^2(t) - 2b(t)\frac{\dot{m}(t)}{m(t)}+
\frac{\dot{m}^2(t)}{4m^2(t)} - \frac{\ddot{m}(t)}{2m(t)}
 - 4b^2(t) - 2\dot{b}(t).                           
\eear
The expressions of $\a(t),\,\b(t)$ and $\g(t)$ in terms of $\e$ and
$\dot{\e}$ are \\[-5mm]

\bear\lb{alfabetagama}
\a(t) &=& \frac{-1}{4\hbar m(t)}\e^2(t),\\   
\b(t) &=&
\frac{-1}{4\hbar}\e(t)\left[2b(t)\e(t)-\dot{\e}(t)+\frac{\dot{m}}{2m}
\e(t)\right],\\                              
\g(t) &= &
-\frac{m(t)}{4\hbar}\left[(2b(t)\e(t)-\dot{\e}(t)+\frac{\dot{m}}{2m}
\e(t)\right]^2.                             
\eear
The above invariants $I_j$ are Hermitian in the space of square integrable
functions on the positive part of the real line which are vanishing at
$x=0$. Such are the wave functions constructed in the next section.

At $0 = dm/dt = dg/dt$  the invariants $I_3(t)$ and $I_-(t) = I_1(t) -
iI_2(t)$ coincide with the corresponding invariants constructed in
\ci{THM81} and at $b= 0 = dm/dt = dg/dt$ they recover those in
\ci{DMM,Gamiz}.  In \ci{Um} one Hermitian invariant ($\sim I_3(t)$) and
its orthonormalized eigenfunctions have been obtained for Hamiltonian
(\ref{H}) with $b=0$ and $c=1$ (i.e.  $2m(t)g(t) = \hbar^2$).  The
invariants $I_j(t)$ will obey the commutation relations of $su(1,1)$
algebra, \\[-5mm]

\be\lb{I_jcomrel}                           
 [I_1(t),I_2(t)]=-iI_3(t),\quad
[I_2(t),I_3(t)]=iI_1(t),\quad [I_3(t),I_1(t)]= iI_2(t),
\ee
if we fix the Wronskian of the solutions $\e(t)$ of the auxiliary
classical oscillator equation (\ref{epsilon}) as \\[-5mm]

\be\lb{Wronskian}                           
\e^*\dot{\e} -
\e\dot{\e}^* = 2i\,\, \longleftrightarrow\,\, \e(t) =
|\e(t)|\exp\left[i\int^t d\tau|\e(\tau)|^{-2}\right].
\ee
The Casimir
operator has the same value as in (\ref{C_2}) (for any time $t$):
$I_3^2(t)-I_1^2(t) - I_2^2(t) = -3/16 + c/4$. The proper initial
conditions which ensure $I_j(0) = L_j$ are ($\O_0 = \O(t\!=\!0)$) \\[-5mm]

\be\lb{cond0}                               
\e(0) = \frac{1}{\sqrt{\O_0}},\,\,
\dot{\e}(0) = i\sqrt{\O_0}; \quad
\dot{b}(0)= \dot{m}(0) = 0,\,\, b(0) = 0.
\ee
With these initial conditions one has:\,\,
(a)  $I_j(t) = U(t)L_jU^\dg(t)$,\,\,
where $U(t)$ is the evolution operator of the general SO, i.e. $I_j(0) =
L_j$ [for other initial conditions $I_j(0)$ is a combination of $L_k$];\,\,
(b) $I_3(0) = H_{\rm SO}(0)/(2\hbar\o_0)$,\,\,
where $H_{\rm SO}(0)$ is the stationary SO Hamiltonian  with mass $m_0$,
frequency $\o_0$ and $g_0 = c\hbar^2/2m_0$.  If in (\ref{cond0}) $b(0)\neq
0$ then $I_3(0) \neq L_3$, but it remains proportional to the initial
Hamiltonian, \\[-5mm]

\be\lb{cond02}                                  
I_3(0) = U^\dg(t)I_3(t)U(t) = H(0)/(2\hbar\o_0),
\ee
where $H(0)$ is the Hamiltonian (\ref{H}) at $t=0$ with $g(0) =
\hbar^2c/2m_0$ (the stationary general SO Hamiltonian). In many papers (see
\ci{Um}, \ci{Angelow} and references therein) solutions to the quantum
(singular) oscillator with varying mass and/or frequency are expressed in
terms of other than $\e(t)$ parameter functions. Regarding the analytic
solutions to the classical equation (\ref{epsilon}) for time-dependent
"frequency" $\O(t)$, see \ci{Angelow,Kamke}. In \ci{CDM} the Heisenberg
operators $U^\dg(t)L_jU(t) = \td{\l}_{jk}(t)L_k$ were constructed for the
initial conditions $b(0)= 0 = \dot{m}(0) = \dot{\o}(0) = \dot{g}(0)$, the
coefficients $\td{\l}_{jk}(t)$ being expressed in terms of a parameter
function $\td{\e}(t)$ which obey a slightly different second order equation.
\vs{5mm}

\section{Wave functions and algebra related coherent states}

Assuming $p = -i\hbar \partial/\partial x$ the wave functions $\Psi(x,t)$
of the general SO (with the constrain (\ref{c})) should obey the
differential (Schr\"odinger) equation \\[-5mm]

\be\lb{SE}                                       
i\hbar\frac{\p}{\p t}\Psi(x,t) =
\frac{1}{2}\left[-\frac{\hbar^2}{m(t)}\frac{\p ^2}{\p x^2} -i\hbar
2b(t)(2x\frac{\p}{\p x} + 1) + m(t)\o^2(t)\, x^2 +
\frac{c\hbar^2}{m(t)}\frac{1}{x^2}\right]\Psi(x,t).
\ee
We shall look for solutions to (\ref{SE}) in the form of
eigenstates of the {\it complex linear combinations} of the constructed
invariants $I_j(t)$.  As in the particular cases of $b=0$
\ci{Um}--\ci{Peak} we consider the collapse free
case $ 1 + 8m(t)g(t)/\hbar^2 = 1 + 4c \geq 0$\, ($c= {\rm const}$)\, and
look for wave functions $\Psi(x,t)$ which are vanishing at $x=0$ (since at
$x\rightarrow 0$ the potential may tend to $\infty$).

We first find the orthonormalized eigenstates $\Psi_n(x,t)$ of the
invariant $I_3(t)$:\\[-5mm]

\be\lb{I_+^n|0>}                                
\Psi_n(x,t) =
\sqrt{\frac{\G(2\k)}{n!\G(2\k+n)}}(I_+(t))^n\Psi_0(x,t),
\ee
where
$I_+(t) = I_1(t) + iI_2(t)$ and $\Psi_0$ is annihilated by $I_-(t) =
I_1(t)- iI_2(t)$:  $I_-(t)\Psi_0(x,t) = 0$. In the above we have
expressed the parameter $c$ in $I_j(t)$ in terms of kappa: $c =
4\k(\k-1) + 3/4$. Since $I_-(t)$ can (as in the particular case of
$\dot{m} = 0 = b$ \ci{DMM}) be casted in the form\\[-5mm]

\be\lb{I_-}                                     
I_-(t) = A(t)^2/2 + c\hbar^2\a(t)/2x^2,
\ee
where $A(t)$ is the invariant boson annihilation operator for the
generalized oscillator ($g = 0$ in (\ref{H})) we put $\Psi_0(x,t) =
\phi(x,t)\psi_0(x,t)$, where $\psi_0(x,t)$ is annihilated by $A(t)$
\ci{MMT2}. Then we easily find $\phi(x,t)$ and construct all
wave functions $\Psi_n(x,t)$ (solutions to eq. (\ref{SE})),\\[-5mm]

\bear\lb{Psi_n}                                  
\Psi_n(x,t) =
\left[2\left(\frac{m(t)}{\hbar\e^2(t)}\right)^{2\k}\frac{n!}{\G(2\k+n)}
\right]^{\frac 12} x^{2\k -\frac 12} \left(\frac{\e^*(t)}{\e(t)}\right)^n\,
\exp\left[-\frac{i}{\hbar}m(t)b(t)x^2\right] \nn\\
\times\exp\left[i\frac{m(t)}{2\hbar\e(t)}\left(\dot{\e}(t)-\frac{\dot{m}(t)}
{2m(t)}\e(t)\right) x^2\right]\,
L^{2\k-1}_n\left(\frac{m(t)}{\hbar|\e(t)|^2}x^2\right),
\eear
where $\e(t)$ obey (\ref{epsilon}) and (\ref{Wronskian}), $\G(z)$ is Gamma
function, $L^d_n(x)$ are generalized Laguerre polynomials \ci{Bateman}.
Since $L^{d}_n(x)$ and $\G(z)$ are defined for Re$\,d > -1$ and Re$\,z >
0$ \ci{Bateman} our functions are square integrable (normalized) for
Re$\,\k > 0$.  It is convenient to also use the Dirac notation
$|\k,\k+n;t\ra$ for the eigenstates of $I_3(t)$ and $|\k,\k+n\ra$ for
$|\k,\k+n;t\!=\!0\ra$,

$$\Psi_n(x,t)
= \la x|\k,\k+n;t\ra,\quad \Psi_n(x,0) = \la x|\k,\k+n\ra.$$
The eigenvalues of $I_3(t)$ are $\k + n$,\\[-5mm]

\be\lb{I3action}                                
I_3(t)|\k,\k+n;t\ra =  (\k+n)|\k,\k+n;t\ra.
\ee
The hermiticity of $I_3$ requires Im\,$\k = 0$, which results in $\k > 0$.
A further restriction on $\k$ follows from the requirement
$\Psi_n(x\!=\!0,t) = 0$, which is satisfied if $$ \k > 1/4.$$ One can easily
show that the latter constrain ensures the hermiticity of $p$ and $x$ in the
space ${\cal H}_\k$ spanned by the wave functions $\Psi_n$ on the positive
part of the real line.  The two values $\k=1/4$ and $\k=3/4$ correspond to
$c=0$ and the states (\ref{Psi_n}) at $\k= 3/4$ coincide (up to constant
factor) with the odd Fock-type states (precisely: squeezed Fock states) of
the general oscillator \ci{MMT1,MMT2}. It is worth noting that the
expression (\ref{Psi_n}) at $\k = 1/4$ formally coincides (up to a constant
factor, which originates from the change of the base coordinate space from
the positive part to the whole real line) with the even Fock-type states of
the general oscillator.  The values $\k \geq 1/2$ are related to $c$ via
$2\k = 1 + (1/2)\sqrt{1+4c}$, $c \geq -1/4$, while for those in the interval
$1/4 < \k < 1/2$ one has $2\k = 1 - (1/2)\sqrt{1+4c}$. Hereafter unless
otherwise stated we consider $2\k = 1 + (1/2)\sqrt{1+4c}$, i.e. the
continuous Bargman index is $\k \geq 1/2$.

 At $t=0$ and (\ref{cond0}) the wave functions $\Psi_n(x,t)$ coincide with
the eigenstates of the initial Hamiltonian $H(0)$ with the energy
$2\hbar\o(\k + n)$.  The expression (\ref{Psi_n}) for $\Psi_n(x,t)$
recovers the corresponding ones for the particular cases, considered
previously \ci{Gamiz,DMM,THM81}.  With (\ref{cond0}) and $b = 0 = dm/dt =
dg/dt$ in (\ref{Psi_n}) the wave functions $\Psi_n(x,t)$ coincide with
those found in \ci{DMM,Gamiz}.  In the particular case of $b=0$ and $c=1$
our $\Psi_n(x,t)$ failed to recover the wave function $\phi_n(q,t)$ of
\ci{Um} (a certain $t$- and $q$-dependent factor is missing in the
expression for $\phi_n(q,t)$).

The family of $\Psi_n(x,t)$ can be used to construct the Green function
$G(x_2,t_2;x_1,t_1)$ for the general SO. From the definition
$$G(x_2,t_2;x_1,t_1) =
\sum_{n=0}^{\infty}\Psi_n(x_2,t_2)\Psi_n^*(x_1,t_1)$$
and by means of the formula for the generating function of the product
$L_n^\a(x)L_n^\a(y)$ of two associate Laguerre polynomials \ci{Bateman}
one can obtain the closed expression\\[-7mm]

\bear\lb{Green}                                    
G(x_2,t_2;x_1,t_1) =
\frac{-i\sqrt{m_1m_2}}{\hbar\r_1\r_2{\rm sin}\g_{12}}
\left(x_1x_2\right)^{1/2}\,\exp\left[\frac{i}{2\hbar}\left(B^*(t_1)x_1^2-
B(t_2)x_2^2\right)\right]\nn \\
\times \exp\left[\frac{i}{2\hbar}{\rm ctan}
\g_{12}\left(\frac{m_1}{\r_1^2}x_1^2 +
\frac{m_2}{\r_2^2}x_2^2\right)\right]\,I_{2\k-1}
\left(\frac{-ix_1x_2\sqrt{m_1m_2}}{\r_1\r_2{\rm sin}\g_{12}}\right),
\eear
where
$$B(t) = m(t)\left[2b(t)-\dot{\r}(t)/\r(t) + \dot{m}(t)/2m(t)\right],\quad
\g_{12} = \int_{t_1}^{t_2}d\tau/|\e(\tau)|^2,$$
\,$\r_i=\r(t_i)\equiv |\e(t_i)|,\,\,m_i=m(t_i),\,\, i = 1,2$,\,\, and
$I_\a(x)$ is the modified Bessel function of the first kind \ci{Bateman}.

 In order to construct more general family of states of the general SO we
note the action of the lowering and raising invariant operators
$I_{\mp}(t)$ on $|\k,\k+n;t\ra$.  From the commutation relations
(\ref{I_jcomrel}) and the construction (\ref{I_+^n|0>}) it follows
that\\[-5mm]

\be\lb{Ipmaction}                                
\left.\begin{tabular}{cc}
$I_-(t)|\k,\k+n;t\ra = \sqrt{n(2\k+n-1)}|\k,\k+n-1;t\ra$, & \\[2mm]
$I_+(t)|\k,\k+n;t\ra = \sqrt{(n+1)(2\k+n)}|\k,\k+n+1;,t\ra$. &
\end{tabular}\right \}
\ee

Noting the analogy of (\ref{I3action}) and (\ref{Ipmaction}) to the case
of the discrete series representation of the algebra $su(1,1)$ \ci{BR} and
the results of papers \ci{Trif9698}\footnote{
     In these articles and in \ci{Trif97} we have used the analytic
     Barut-Girardello representation on the complex plane in diagonalizing
     the general element of $su^C(1,1)$ (the discrete series $D^+(k)$).
     For the same diagonalizations the analytic representation on the unit
     disk was applied by Brif (first paper of \ci{Brif}).}
we construct the following family of solutions to the Schr\"odinger
equation (\ref{SE}):\\[-5mm]

\be\lb{|zuvw>}                                    
|z,u,v,w;\k,t\ra = N \sum_{n=0}^{\infty}
a_n(z,u,v,w,\k)|\k,\k+n;t\ra,
\ee
where $N$ is the normalization factor,\, $z,u,v$ and $w$ are complex
parameters and \\[-5mm]

\be\lb{a_n}                                      
a_n(z,u,v,w,\k) = \left(-\frac{l+w}{2u}\right)^n
\sqrt{\frac{(2\k)_n}{n!}}\,
_2F_1\left(\k+\frac{z}{l},-n;2\k;\frac{2l}{l+w}\right).
\ee
Here $l = \sqrt{w^2-4uv}$, $(a)_n$ is Pochhammer symbol and
$_2F_1(a,b;c;z)$ is the Gauss hypergeometric function \ci{Bateman}.

Using the actions (\ref{I3action}), (\ref{Ipmaction}) and the relation
(formula 2.8.40 of \ci{Bateman})
$$
(c-2b+bz-az)\,_2F_1(a,b;c;z) + b(1-z)\,_2F_1(a,b+1;c;z) +
 (b-c)\,_2F_1(a,b-1;c;z) = 0 $$
one can verify (after somewhat tedious calculations) that the above
solutions are {\it eigenstates of the complex combination} of the three
invariant operators $I_j(t)$,\\[-5mm]

\be\lb{zuvwstates}                                
\left[uI_-(t) + vI_+(t)+ wI_3(t)\right]\,|z,u,v,w;\k,t\ra =
z|z,u,v,w;\k,t\ra.
\ee
Note that under the initial conditions (\ref{cond0}) we have $[uL_- +
vL_+ +wL_3]\, |z,u,v,w;\k,0\ra = z|z,u,v,w;\k,0\ra$, since under that
conditions $I_{\pm}(0) = L_{\pm} \equiv L_1\pm iL_2$ and $I_3(0) =
L_3$.

The states (\ref{|zuvw>}) are normalized but not orthogonal.  Their
scalar product for different parameters $z,u,v,w$ and
$z^\pr,u^\pr,v^\pr,w^\pr$ but the same real $\k$ can be obtained (by
making use of formula 2.5.1.12 of \ci{Bateman}) in the form\\[-7mm]

\bear\lb{<z|z^pr>}                                
\la t,\k;w,v,u,z\,|\,z^\pr,u^\pr,v^\pr,w^\pr;\k,t\ra  =
N^\pr\,N\,(1+s)^{z^*/l^* + z^\pr/l}
(1+s-s\z^*)^{-\k-z^*/l^*}\nn \\
\times(1+s-s\z^\pr)^{-\k-z^\pr/l}\,_2F_1\left(\k+\frac{z^*}{l^*},\k
+\frac{z^\pr}{l^\pr};2\k; \frac{-s\z^*\z^\pr}{(1+s -
s\z^*)(1+s-s\z^\pr)}\right),
\eear
where $s =
-(w^*+l^*)(w^\pr+l^\pr)/(4u^*u^\pr)$, $\z = 2l/(w+l)$, $\z^\pr =
2l^\pr/(w^\pr +l^\pr)$.  The normalization factor $N$ in (\ref{|zuvw>})
is $N = [\la t,\k;w,v,u,z|z,u,v,w;\k,t\ra]^{-1/2} =
N(s,\z,\k+z/l,\k)$,\\[-7mm]

\bear\lb{N}                                        
N^{-2}(s,\z,\k+z/l,\k)  =
(1+s)^{2{\rm Re}(z/l)}\left|(1-s-s\z)^{-(\k+z/l)}\right|^2\nn\\
   \times
   _2F_1\left(\k+\frac{z^*}{l^*},\k+\frac{z}{l};2\k;\frac{-s|\z|^2}
                {|1+s-s\z|^2}\right).
\eear
 The expressions (\ref{<z|z^pr>}) and (\ref{N}) are correct if the
parameter $s$ is small \ci{Bateman}, $|s| < 1$, i.e.

$$|w+\sqrt{w^2-4uv}| <  2|u|.$$
The limit $l=0$ in formulas  (\ref{|zuvw>}) - (\ref{N}) can  safely be
taken.

In the above we have considered $u \neq 0$.  For the case $u=0$ (where
(\ref{N}) is meaningless) we treat the eigenvalue equation
 (\ref{zuvwstates}) separately and find the following normalized solutions
 $|z,u\!=\!0,v,w;\k,t\ra = |z_m,v,w;\k,t\ra$\\[-5mm]

\be\lb{u=0}                                      
|z_m,v,w;\k,t\ra = C_m(v,w)\,\sum_{n=0}^{\infty}\frac{(-v/w)^n}{n!}\,
\sqrt{(m+n)!(2\k)_{m+n}}\,|\k,\k+m+n;t\ra,
\ee
where $z_m$ is the eigenvalue of $vI_-(t) + wI_3(t)$,
$z_m = w(\k+m)$, $m = 0,1,2,\ldots$, and the normalization factor $C_m$
reads ($(1)_m = m!$)\\[-5mm]

\be\lb{C_m}                                       
C_m (v,w) = \left[(1)_{m}(2\k)_m\,\,
_2F_1\left(m+1,2\k+m;1;|v/w|^2\right)\right]^{-\frac
12}.
\ee

 Let us note some important particular cases of the above states.  The
value $v=0$ in (\ref{u=0}) is admissible and it reproduces the eigenstates
$|\k,\k+m;t\ra$ of $I_3(t)$, the wave functions of which are
 given in (\ref{Psi_n}).  The values $v=0=w$ (and $u=1$) in (\ref{|zuvw>})
are also admissible and in this way we obtain the eigenstates
$|z;\k,t\ra = |z,u\!=\!1,v\!=\!0,w\!=\!0;\k,t\ra$ of the lowering
 invariant $I_-(t)$ as a particular case of (\ref{zuvwstates}): \\[-5mm]

\be\lb{|z,t>}                                    
|z;\k,t\ra = \left[_0F_1(2\k;|z|^2)\right]^{-\frac
12} \sum_{n=0}^{\infty}\frac{z^n}{\sqrt{n!(2\k)_n}}|\k,\k+n;t\ra.
\ee
Using the generating function for the Laguerre polynomials $\sum_n^\infty
zL_n^\a(x)/\G(\a+n+1)) = (xz)^{-\a/2}e^zJ_\a(2\sqrt{xz})$ \ci{Bateman} we
obtain the normalized wave functions $\Psi_z(x,t) = \la x|z;\k,t\ra$
in the form\\[-7mm]

\bear\lb{Psi_z}                                
\Psi_z(x,t) =
\left[_0F_1(2\k;|z|^2)\right]^{-\frac 12}
z^{-\k+1/2}\left[\frac{2m(t)}{\hbar\e^2(t)}x\right]^{\frac 12}
J_{2\k-1}\left( \frac{2}{\e(t)}\sqrt{\frac{m(t)}{\hbar}}\,x\right)\nn\\
\times\exp\left[-\frac{im(t)}{2\hbar} \left(2b(t)+
\frac{\dot{m}(t)}{m(t)} - \frac{\dot{\e}(t)}{\e(t)}\right)x^2 +
z\frac{\e^*(t}{\e(t)}\right],
\eear
where $J_\a(x)$ is the Bessel function \ci{Bateman}. At $b = 0$ and
$\dot{m}=0$  (also $z=\a^2/2$) our $\Psi_z(x,t)$ recover the corresponding
wave functions for SO in \ci{Gamiz,DMM}.
One can check that this continuous family is
overcomplete in the space ${\cal H}_\k$ spanned by the orthonormalized
wave functions $\Psi_n(x,t)$,\\[-5mm]

\be\lb{overcompl1}                               
\int d^2z\,f_1(|z|)|z;\k,t\ra\la t,\k;z| =
\sum_{n=0}^{\infty}|\k;\k+n\ra\la n+\k,\k| \equiv 1_\k,
\ee
where the weight function is $$f_1(|z|) = \frac{2}{\pi}
K_{2\k-1}(2|z|)\,I_{2\k-1}(2|z|),$$ $I_d(x)$ and $K_d(x)$ being the
modified Bessel functions of the first and the third kind correspondingly.
Therefore the states (\ref{|z,t>}) can be considered as an extension of
the Barut-Girardello CS \ci{BR} to the case of continuous representations
realized by the invariants $I_j(t)$.

Another particular case of the general family $|z,u,v,w;\k,t\ra$ to
be noted is that of \\[-5mm]

\be\lb{uvwconstrain}                             
u = {\rm cosh}^2r,\quad v = {\rm sinh}^2r\,e^{2i\t},\quad w= {\rm
sinh}(2r)\,e^{i\t},\quad r\geq 0.
\ee
Under this choice  $l^2 = w^2-4uv = 0$ and the solutions
$|z,u,v,w;\k,t\ra$ take the form of the Klauder-Perelomov $SU(1,1)$
group-related CS (see \ci{KS} and references therein)

$$|\z;\psi_0,t\ra = S(\z,t)|\psi_0\ra,\quad S(\z,t) =\exp\left(\z
I_+(t) - \z^*I_-(t)\right),\,\z = re^{i\t},$$
with the "fiducial" vector $|\psi_0\ra = |z;\k,t\ra$, where $|z;\k,t\ra$
is  the state (\ref{|z,t>}).  This parameter identification is based on
the BCH formula

$$S(\z)I_-S^\dg(\z)={\rm cosh}^2r\,I_- + e^{-2i\t}{\rm sinh}^2r\,I_+ +
 e^{-i\t}{\rm sinh}2r\, I_3.$$
If furthermore $z=0$ one gets the analogues of the $SU(1,1)$ group-related
CS with maximal symmetry for the nonstationary general SO in the
continuous representation, generated by the invariants $I_j(t)$ ($\xi =
-{\rm tanh}|\z|\exp[-i\t]$, $|\xi| <1$),\\[-5mm]

\be\lb{|zeta,k>}                                 
|\xi;\k,t\ra := S(\z,t)|\k,\k;t\ra =
(1-|\xi|^2)^{\k}
\sum_{n=0}^{\infty}\sqrt{\frac{(2\k)_n}{n!}}\xi^n
 |\k,\k+n;t\ra.
\ee
The wave function $\Psi_{\xi}(x,t)=\la x|\xi;\k,t\ra$ can be
obtained by means of the generating function for the Laguerre polynomials
$\sum_n \xi^n L_{n}^{\a}(x) = (1-\xi)^{-\a-1}\exp[x\xi/(\xi-1)]$
\ci{Bateman},\\[-5mm]

\bear\lb{Psi_xi}                              
\Psi_{\xi}(x,t)\, = \,
 (1-|\xi|^2)^{\k}\sqrt{\frac{2}{\G(2\k)}}\left(\frac{m(t)}
{\hbar\e^2(t)}\right)^{\k}
\left(1-\xi\frac{\e^*(t)}{\e(t)}\right)^{-2\k}x^{\k}\nn \\
\times \exp\left[\frac{m(t)}{\hbar\e(t)}\frac{x^2\xi}{\xi
\e^*(t)- \e(t)}\right]
\exp\left[-i\frac{m(t)}{2\hbar}\left(2b(t)+\frac{\dot{m}(t)}{2m(t)}-
\frac{\dot{\e}(t)}{\e(t)}\right)x^2\right].
\eear
The continuous family of states $\Psi_\xi(x,t)$ can also resolve the unity
$1_\k$ in the space ${\cal H}_\k$, \\[-5mm]

\be\lb{overcompl2}                             
1_k = \int_{\xi\in I\!\!D}d^2\xi f_2(|\xi|)|\xi;\k,t\ra\la t,\k;\xi|,
\quad f_2(|\xi|) = \frac{2\k-1}{\pi}(1-|\xi|^2)^{-2},
\ee
where $I\!\!D$ is the unit disk in the complex plane.

At $b=0$ and $\dot{m} = 0 = \dot{g}$ and under substitution $z = \a^2/2$
our wave functions $\Psi_z(x,t)$ and $\Psi_\xi(x,t)$ recover those in
papers \ci{Gamiz,DMM} (reproduced also in \ci{DMR}).  Moreover, it can be
verified that at $\k=1/4,3/4$ the wave functions $\la x|z,u,v,w;\k,t\ra$
recover (up to a constant factors) the even and odd wave functions for the
general oscillator.

Next we shall briefly discuss the intelligent \ci{Aragone,Trif94} and the
squeezing \ci{Walls,Trif94} properties of states $|z,u,v,w;\k,t\ra$ [the
term intelligent states was introduced in \ci{Aragone} for the states
which minimize the Heisenberg relation for the spin components].  For real
$w$ and $v=u^*$ the operator $uI_-(t) + u^*I_+(t) + wI_3(t)$ is Hermitian,
therefore \ci{Trif97} the states $|z,u,u^*,w\!=\!w^*;\k,t\ra$ minimize the
Robertson inequality \ci{Rob,Trif97} for the three observables $I_j$
($\vec{I} = (I_1,I_2,I_3)$) :\\[-5mm]

\be\lb{Rob-rel}                              
\det\s(\vec{I}) = \det C(\vec{I}),
\ee
where $\s$ is the uncertainty matrix, $\s = \{\s_{ij}\}$,

$$\s_{ij} = \la I_iI_j + I_jI_i\ra/2 - \la I_i\ra\la I_j\ra \equiv \D
I_iI_j,$$
and $C_{ij} = -i\la[I_i,I_j]\ra/2$.  The quantity $\s_{ii} = \D I_iI_i =
\D^2 I_i$ is called the variance of $I_i$. The Robertson relation for
two observables is known as Schr\"odinger one.  The large family
$|z,u,v,w;\k,t\ra$ contains the full sets of $I_i$-$I_j$ generalized
intelligent states, which are defined \ci{Trif94} as states minimizing the
Schr\"odinger inequality \\[-5mm]

\be\lb{Sch-rel}                              
\D^2 I_i \D^2 I_j - (\D I_iI_j)^2 \geq |\la[I_i,I_j]\ra|^2/4,
\ee
and therefore could also be called Schr\"odinger minimum uncertainty states
or Schr\"odinger intelligent states.  For example the states

$$|z,u,v,w\!=\!0;\k,t\ra \equiv |z,u,v;\k,t\ra$$
are $I_1$-$I_2$ Schr\"odinger intelligent states:  the three second moments
of $I_1$ and $I_2$ in $|z,u,v;\k,t\ra$ are \\[-5mm]

\be\lb{2moments}                                  
\D^2 I_1 = \frac{1}{2}\frac{|u-v|^2}{|u|^2-|v|^2}\la I_3\ra,\quad
\D^2 I_2 = \frac{1}{2}\frac{|u+v|^2}{|u|^2-|v|^2}\la I_3\ra,\quad
\D I_1I_2 = \frac{{\rm Im}(u^*v)}{|u|^2-|v|^2}\la I_3\ra,
\ee
and one can readily check that they minimize (\ref{Sch-rel}). The
covariances of $I_1$ and $I_3$ and of $I_2$ and $I_3$ in $|z,u,v;\k,t\ra$
read simply \\[-5mm]

\be\lb{s13,s23}                                  
\D I_1I_3 = {\rm Re}z/2,\quad \D I_2I_3 = -{\rm Im}z/2.
\ee
The mean values $\la (I_3(t))^m\ra$, $m=1,2,\ld$, in the general state
$|z,u,v,w;k,t\ra$ can be calculated by means of the analytic
formula\\[-5mm]

\be\lb{<I_3>}                                    
\la (I_3)^m\ra = N^2(s,\z,\k+z/l,\k)\, [\k + s\frac{\partial}{\partial
s}]^m \,N^{-2}( s,\z,\k+z/l,\k),
\ee
$N(s,\z,\k+z/l,\k)$ being defined in eq. (\ref{N}). If $w=0$ then the
condition $|s|<1$ results in $|v| < |u|$, which coincides with the
normalizability condition for the eigenstates $|z,u,v;k\ra$ of $uK_- +
vK_+$ \ci{Trif94}, where $K_{\pm}$ are Weyl operators of $su(1,1)$ in the
discrete series $D^+(k)$.

It is remarkable that the wave functions $\Psi_\xi(x,t)$ minimize Robertson
relation (\ref{Rob-rel}) for the three invariants $I_1(t),\,I_2(t),\,I_3(t)$
and Schr\"odinger inequality for all three pairs $I_i,\,I_j$ {\it
simultaneously}. Therefore $\Psi_\xi(x,t)$ minimize the third- and the
second-order characteristic uncertainty relations \ci{TD98} simultaneously.
[The Robertson inequality for $n$ observables relates the $n$th order
characteristic coefficients of matrices $\s$ and $C$. It was established
recently \ci{TD98} that similar inequalities hold for all the other
characteristic coefficients.]  These intelligent properties of
$\Psi_\xi(x,t)$ can be directly checked by calculation of the first and
second moments of $I_j(t)$, \\[-5mm]

\be\lb{<I_j>}                                    
\la I_1\ra = 2\k\frac{{\rm Re}\xi}{1-|\xi|^2},\quad
\la I_2\ra = -2\k\frac{{\rm Im}\xi}{1-|\xi|^2},\quad
\la I_3\ra = \k\frac{1+|\xi|^2}{1-|\xi|^2},\quad
\ee

\be\lb{VarI_j>}                                 
\D^2 I_1 = \frac{\k}{2}\frac{|1+\xi^2|^2}{(1-|\xi|^2)^2},\quad
\D^2 I_2 = \frac{\k}{2}\frac{|1-\xi^2|^2}{(1-|\xi|^2)^2},\quad
\D^2 I_3 = 2\k\frac{|\xi|^2}{(1-|\xi|^2)^2},\quad
\ee

\be\lb{CoVarI_j>}                               
\D I_1I_2 = -2\k\frac{{\rm Re}\xi{\rm Im}\xi}{(1-|\xi|^2)^2},\quad
\D I_1I_3 = \k{\rm Re}\xi\frac{1+\xi^2}{(1-|\xi|^2)^2},\quad
\D I_2 I_3 = -\k{\rm Im}\xi\frac{1+|\xi|^2}{(1-|\xi|^2)^2},\quad
\ee
So $\Psi_\xi$ are states with {\it maximal characteristic intelligency}.

The analysis, similar to that for the states $|z,u,v;k\ra$
\ci{Trif94,Trif9698}, shows that the variance of $I_1$ ($I_2$) can tend to
zero when $v\rightarrow u$ ($v\rightarrow -u$).  Therefore the SO states
$|z,u,v;\k,t\ra$ are {\it ideal $I_1$-$I_2$ squeezed states}
\ci{Trif9698}. $I_1$ ($I_2$) squeezing can also occur in states
$|u,v\!=\!u,w\!\neq\!0;\k,t\ra$ ($|u,v\!=\!-\!u,w\!\neq\!0;\k,t\ra$),
which minimize the Schr\"odinger relation for $I_1$ and $I_3$ ($I_2$ and
$I_3$).

One sees that the above moments of $I_{j}(t)$ in $|z,u,v,w;\k,t\ra$ are
time independent in accordance with the fact that $I_j(t)$ are
exact invariants of the system (\ref{H}). Under the initial conditions
(\ref{cond0}) all the moments of $I_j(t)$ in $|z,u,v,w;\k,t\ra$ coincides
with those of $L_j$, equation (\ref{L_j}), in $|z,u,v,w;\k,t\!=\!0\ra
\equiv |z,u,v,w;\k\ra$.  The three second moments of $L_j$ in
$|z,u,v;\k\ra$ are given by the same formulas (\ref{2moments}) with $I_j$
replaced by $L_j$. So the states $|z,u,v;\k\ra $ are $L_1$-$L_2$ ideal
squeezed state.  The time evolution of the moments of $L_j$ can be
obtained by expressing $L_j$ in terms of the invariants $I_j(t)$: $L_j =
\L^{-1}_{jk}(t)I_k(t)$. The coefficients $\l_{jk}(t)$ can be easily
calculated from (\ref{I_1})--(\ref{I_3}). The matrix $\L(t)$ takes the
form

\be\lb{Lambda}                                     
\Lambda = 8\hbar^2\left(
\matrix{ \frac{1}{4\hbar}
\left(\frac{{\rm Re}\g}{m_0\o_0}-m_0\o_0{\rm Re}\a\right)
\qquad
\frac{1}{4\hbar}\left(\frac{{\rm Re}\g}{m_0\o_0} - {\rm Re}\b\right)
\qquad
 \frac{1}{4\hbar} m_0\o_0{\rm Re}\a \\[3mm] \cr
\frac{1}{4\hbar}\left(\frac{{\rm Re}\g}{m_0\o_0} - m_0\o_0{\rm Im}\a\right)
\qquad
\frac{1}{4\hbar}\left( \frac{{\rm Re}\g}{m_0\o_0} - {\rm Im}\b\right)
\qquad
\frac{1}{4\hbar}m_0\o_0{\rm Im}\a   \\[3mm]   \cr
\frac{{\rm Im}(\b\g)}{m_0\o_0}-m_0\o_0{\rm Im}(\a\b^*)
\qquad
 \frac{{\rm Im}(\b\g)}{m_0\o_0} - {\rm Im} (\a\g^*)
\qquad
m_0\o_0{\rm Im}(\a\b^*) }   \right)
\ee

Then the second moments $\s_{ij}(\vec{L})$ of $L_j$ in any state are
simply related to those of $I_j(t)$ \ci{Trif9698,Trif97}:

\be\lb{sig-sig^pr}                                 
\s_{ij}(\vec{L}) = \l_{in}(t)\s_{nm}(\vec{I})\l_{jm}(t).
\ee

\section{Concluding remarks}

We have constructed three linearly independent invariants $I_j(t)$ for the
general nonstationary SO (\ref{H}) with (\ref{c}) (the $su(1,1)$
SO) and diagonalized their general complex combination $(u+v)I_1 +
i(v-u)I_2 + wI_3$. The initial conditions under which $I_j(0)$ coincide
with the familiar $su(1,1)$ operators $L_j$, eq. (\ref{L_j}), are those
of eq. (\ref{cond0}). In several particular cases the closed expressions
for the wave functions in coordinate representation is provided.  The
general family of the diagonalizing states $|z,u,v,w;\k,t\ra$, eq.
(\ref{|zuvw>}), recovers all states previously constructed and contains
all states which minimize the Robertson uncertainty relation for three
observables $I_{1,2,3}(t)$ and all states which minimize Schr\"odinger
relation for any pair of observables $I_j(t),\,I_k(t)$. It is established
that the states $\Psi_\xi(x,t)$, eq.  (\ref{Psi_xi}), are with maximal
intelligency, minimizing the Robertson relation for the
three invariants $I_j(t)$ and the Schr\"odinger inequality for all pairs
$I_j(t),\,I_k(t)$ {\it simultaneously}.  Therefore $\Psi_\xi(x,t)$
minimize the third- and the second-order characteristic uncertainty
relations, established in \ci{TD98}.

If the singular perturbation in (\ref{H}) is switched off ($g = 0
\rightarrow \k=1/4,\,3/4$), the wave functions $\la
x|z,u,v,w;\k\!=\!1/4,3/4,t\ra$ can reproduce (up to constant factors) the
corresponding time-evolved even and odd states for the general quadratic
Hamiltonian \ci{Trif9698,Brif}, various particular cases of which are
widely discussed in the literature \ci{Walls,Angelow}.

Since the invariants $I_j(t)$ close the Lie algebra $su(1,1)$ in the
continuous representation (\ref{C_2}) the family of the diagonalizing
states $|z,u,v,w;\k,t\ra$, eq.  (\ref{|zuvw>}), may be called $su^C(1,1)$
algebra-related coherent states \ci{Trif9698} for the SO. Many of the
formulae concerning the SO states $|z,u,v,w;\k,t\ra$  (first and second
moments formulae, scalar products, resolution of unity) remain valid for
the discrete series $D^+(k)$ of $su(1,1)$ by fixing the continuous
parameter kappa $\k$ equal to the discrete Bargman index
$k=1/2,1,3/2,\ldots$ or $k=1/4,3/4$. It is worth recalling here that
$D^+(k)$ of $SU(1,1)$, $k= 1/2,1,3/2,\ldots$, has the important
realization in the space of states of two mode boson/photon systems with
fixed difference of numbers of bosons/photons in the two modes. The
$su^C(1,1)$ algebra eigenstates for $D^+(k)$ have been studied in detail in
\ci{Trif94,Trif9698,Brif}.  \vs{5mm}

{\bf Acknowledgement}. This work is partially supported by the Bulgarian
Science Foundation grants No F-559 and F-644. The author thanks V.V.
Dodonov for valuable remarks.

\newpage
\vs{5mm}


\begin{thebibliography}{99}

\bibitem{DMR}         
    Dodonov V V, Man'ko V I and Rosa L 1998 Phys. Rev. A
    {\bf57}  2851 {\sm (and references therein)}
   "Quantum singular oscillator as a model of a two-ion trap:  An
   amplification of transition probabilities due to small time variations 
   of the binding potential". 
  
\bibitem{Um}         
   Um C I, Shin S M, Yeon K H and George T F 1998 Phys. Rev. A {\bf58}
   1574
  "Exact wave function of harmonic plus inverse harmonic
   potential with time dependent mass and frequency".

\bibitem{Kaushal}    
     Kaushal R S and Parashar D 1997 Phys. Rev. A {\bf55} 2610
   "Time dependent harmonic plus inverse harmonic potential in quantum
    mechanics".

\bibitem{Simon}     
    Simon B D, Lee P A and  Altshuler B L 1994 Phys. Rev. Lett. {\bf72} 64
  "Matrix models, One-Dimensional Fermions and Quantum Chaos".

\bibitem{DMZ}        
    Dodonov V V, Man'ko V I and Zhivotchenko D V 1993
    N. Cimento B {\bf108} 1349
   "Quasienergies and chaotic behaviors of periodically delta-kicked
    quantum singular oscillator",

\bibitem{Cheng}     
    Cheng B K and  Chang F T 1987 J. Phys. A: Math. Gen. {\bf 20} 3771
   "An exact propagator for a time-dependent harmonic oscillator with a
    time-dependent inverse square potential".

\bibitem{CDM}        
    Chumakov S M, Dodonov V V and Man'ko V I 1986
    J. Phys. A: Math. Gen. {\bf 19} 3229
    "Correlation functions of the non-stationary quantum singular
     oscillator".

\bibitem{Korsch}    
  Kaushal R S and  Korsch H J  1981  J. Math. Phys. {\bf22} 1904

\bibitem{THM81}     
    Trifonov D, Hristova P and Milanov G 1981  Annual Higher Ped. Inst. B
    {\bf5} 49
    "Parametric excitation of the generalized singular oscillator".

\bibitem{Ray}       
      Ray J R and  Reid J L 1979   Phys. Lett. A {\bf 74} 23
    "Exact time-dependent invariants  for $N$-dimensional systems".

\bibitem{DMM}       
  Dodonov V V, Malkin I A and Man'ko V I 1974  Physica {\bf 72} 597\\
  Malkin I A and Man'ko V I. 1979 {\it Dynamical symmetries and coherent
  states of quantum systems} (Moscow: Nauka)

\bibitem{Calogero}    
  Calogero F 1969  J. Math. Phys. {\bf10} 2191\\
  Calogero F 1971  J. Math. Phys. {\bf12} 419
  "Solution of the one-dimensional N-body problem with quadratic
   and/or inverse quadratic pair potential".

\bibitem{Gamiz}     
  Gamiz P, Gerardi A, Marshioro C, Presutti E and Scacciatelli E
   1971  J. Math. Phys. {\bf22} 2040
   "Exact solution of time-dependent quantal harmonic oscillator
   with singular perturbation".

\bibitem{Peak}      
   Peak D and Inomata A  1969  J. Math. Phys. {\bf10} 1422
   "Summation over Feynman histories in polar coordinates".

\bibitem{BR}        
   Barut A O and Raczka R. 1977 {\it Theory of Group Representations and
   Applications} (Warszawa: Polish Scientific Publishers)
  (Russian transl. 1980 (Moskva: Nauka))

\bibitem{Lewis}     
    Lewis H R and Riesenfeld W B 1969   J. Math. Phys. {\bf10} 1458

\bibitem{MMT1}      
   Malkin I A,  Man'ko V I and Trifonov D A 1969  Phys. Lett. A {\bf
   30} 414\\
   Malkin I A,  Man'ko V I and Trifonov D A 1970   Phys. Rev. D{\bf
   2} 1371
   "Coherent States and Transition Probabilities in a Time-Dependent
    Electromagnetic Field".

\bibitem{MMT2}      
   Malkin I A,  Man'ko V I and Trifonov D A 1971 Nuovo Cimento A {\bf
   4} 773
   "Dynamical symmetry of nonstationary systems"\\
   Malkin I A,  Man'ko V I and Trifonov D A 1973 J. Math. Phys. {\bf
   14} 573
   "Linear adiabatic invariants and coherent states"
   
\bibitem{KS}        
   Klauder J R  and  Skagerstam B S 1985  {\it Coherent States}
        (Singapore: World Scientific)

\bibitem{Walls}     
   Walls D F 1983 Nature, {\bf 306} 141
   "Squeezed states of light";\\
   Loudon R and Knight P 1987  J. Mod. Opt. {\bf 34} 709 
   "Squeezed light".\\
   Man'ko V I 1992 in {\it Quantum Measurements in Optics}, Eds. P.
     Tombesi P and Walls D (New York: Plenum Press) (NATO ASI Series B,
     v. 282)
   "Time-Dependent Invariants and Nonclassical Light".

\bibitem{Trif94}       
   Trifonov D A 1994 J. Math. Phys. {\bf 35} 2297 
   "Generalized Intelligent States and Squeezing".\\
   Trifonov D A 1993  Generalized Intelligent States and $SU(2)$ and
   $SU(1,1)$ Squeezing {\it Preprint} INRNE-TH-93/4 (May 1993)

\bibitem{Aragone}     
   Aragone C, Chalbaud E and Salamo S 1976  J. Math. Phys.{\bf 17} 1963

\bibitem{Rob}         
   Robertson  H P 1934   Phys. Rev. {\bf46} 794 \\
   Schr\"odinger E 1930  {\it Sitz. der Preuss. Acad. Wiss.
   (Phys.-Math. Klasse)} (Berlin) p 296

\bibitem{Hillery}        
     Bergou J A,  Hillery M and Yu D  1991  Phys. Rev. A {\bf43} 515 
     "Minimum uncertainty states for amplitude-squared squeezing:
      Hermite polynomial states".\\
     Nieto M M and Truax D R 1993   Phys. Rev. Lett. {\bf71} 2843
     "Squeezed States for General Systems".

\bibitem{Trif9698}       
     Trifonov D A  1996 Algebraic coherent states and
     squeezing  {\it E-print} quant-ph/9609001 \\
     Trifonov D A  1998  Phys. Scripta {\bf 58} 246
     "On the squeezed states for n observables".

\bibitem{Brif}           
     Brif C  1997  Int. J. Theor. Phys. {\bf 36} 1677 
     "$SU(2)$ and $SU(1,1)$ Algebra Eigenstates: A Unified Analytic
      Approach to Coherent and Intelligent States".\\
     Brif C 1996 Ann. Phys. {\bf 251} 180
     "Two-Photon Algebra Eigenstates. A unified Approach to Squeezing."

\bibitem{Trif98}         
     Trifonov D A 1998  Canonical equivalence of quantum systems,
      multimode squeezed states and Robertson relation
      {\it E-print} quant-ph/9801015

\bibitem{Yeon}           
     Yeon K H, Kim D H, Um C I, George T F and  Randey L N 1997  Phys. Rev.
     A {\bf55} 4023\\
     Yeon K Y, Kim H J, Um C I, George T F
     and  Randey L N  1994  Phys. Rev. A {\bf50} 1035

\bibitem{Angelow}        
     Angelow A  1998  Physica A {\bf256} 485\\
     "Light propagation in nonlinear waveguide and classical
      two-dimensional oscillator";\\
     Angelow A  Frequency Switching of Quantum Harmonic Oscillator with
          time-dependent frequency  {\it E-print} quant-ph/9803004

\bibitem{Kamke}        
     Kamke E\, 1959 {\it Differentialgleichungen, L\"osungsmethoden
     und L\"osungen} (Gew\"ohnliche differentialgleichungen)
     (Leipzig: Akademische Verlagsgesellschaft) v. 1

\bibitem{Bateman}        
     Bateman H and Erd\'elyi A 1953 {\it Higher transcendental
      functions} (New York: McGraw-Hill) (Russian
      transl. 1973 (Moskva: Nauka)

\bibitem{Trif97}         
     Trifonov D A 1997  J. Phys. A: Math. Gen. {\bf 30} 5941
     "Robertson Intelligent States".

\bibitem{TD98}         
    Trifonov D A and Donev S G 1998  J. Phys. A: Math. Gen. {\bf 31} 8041
    "Characteristic Uncertainty Relations".

\end{thebibliography}
\end{document}